\documentstyle[12pt]{article}
\newcommand{\beq}{\begin{equation}}
\newcommand{\bea}{\begin{eqnarray}}
\newcommand{\eeq}{\end{equation}}
\newcommand{\eea}{\end{eqnarray}}
\newcommand{\bdm}{\begin{displaymath}}
\newcommand{\edm}{\end{displaymath}}
\newcommand{\eps}{\epsilon}

\newcommand{\qplus }  {q^{2<\eps_\omega,\rho>}}
\newcommand{\qmins }  {q^{-2<\eps_\omega,\rho>}}

\newcommand{\build}[3]{\mathrel{\mathop{\kern 0pt#1}\limits_{#2}^{#3} }}

\catcode`\@=11
\def\numberbysection{\@addtoreset{equation}{section}
        \def\theequation{\thesection.\arabic{equation}}}
\numberbysection

\begin{document}
\setcounter{page}{0}[M\topmargin 0pt \oddsidemargin 5mm
\renewcommand{\thefootnote}{\fnsymbol{footnote}} \newpage
\setcounter{page}{0}
\begin{titlepage}
\begin{flushright}
\vspace{-1cm}
PUPT 1536
\end{flushright}
\begin{center}
{\Large\bf  Exact solution of A-D Temperley-Lieb Models}\\
\vspace{1cm}
{\large    R. K\"{O}BERLE$^{1\dagger}$}\footnote{
On leave of abscence from Instituto de F\'{\i}sica
 de S\~ao Carlos,
              Universidade de S\~ao Paulo,\\
               Caixa Postal 369, S\~ao Carlos 13560-970 , Brasil.  \\
$^{\dagger}\;\;\;$Supported in part by CNPq-BRASIL.}\\
             and \\
 {\large{         A. LIMA - SANTOS$^{2\dagger}$}}\\
$^1$Dept. of Physics, Princeton University, Princeton, NJ 08544\\
\vspace{-.3cm}
\begin{small}and \end{small}\\
\vspace{-.3cm} NEC Research Institute, 4 Independence Way, Princeton, NJ 08540
\\
{\small e-mail: roland@puhep1.princeton.edu} \\
\vspace{.2cm}
$^2$Departamento de F\'{\i}sica, Universidade
Federal de S\~ao Carlos\\
             Caixa Postal 676, 13569-905 S\~ao Carlos, Brasil\\
\vspace{.5cm}
     \begin{abstract}

 We solve for the spectrum of quantum spin chains based on representations of
the Temperley-Lieb algebra associated with the quantum groups ${\cal U}_q(X_n)$
for $X_n = A_1,B_n,C_n$ and $D_n$.
We employ a generalization of the coordinate Bethe-Ansatz developed previously
for the deformed biquadratic spin one chain.
As expected, all these models have equivalent spectra, i.e. they differ only
in the degeneracy of their eigenvalues. This is true for finite length and
open boundary conditions. For periodic boundary conditions the spectra of
the lower dimensional representations are containded entirely in the higher
dimensional ones. The Bethe states are highest weight states of the
quantum group, except  for some states with energy zero.
\end{abstract}
%
 \end{center}
\end{titlepage}
\newpage

\section{Introduction}

\label{sect.1}

The recent interplay between the field of solvable two-dimensional lattice (
or quantum spin chain ) models and quantum groups, has generated a lot of
interesting results. One particular way of building models, which are
quantum group invariant, uses the Temperley-Lieb (TL) algebra\cite{TL},
satisfied by the Hamitonian density $U_k$:

\[
U_k^2 = \sqrt{Q}U_k,\;\;U_k U_{k\pm 1} U_k=U_k,\;\;
\]
\begin{equation}
[U_k,U_l]=0,\;\;|k-l|\geq 2.
\end{equation}

The Hamiltonian is now given by the following sum over $N$ sites:
\begin{equation}
H(q) = \sum_{k=1}^N U_k.
\end{equation}

The paper is organized as follows. In section 2, we describe the
representations of the TL algebra, constructed as projectors on total spin
zero of two neighbouring spins. In section 3, we discuss a modified
coordinate BA and show the reasons, why the techniques develloped for the
spin 1/2 XXZ model don't work here. Section 3 contains the algebraic details
of the computation and section 4 is reserved for the conclusions.

\section{Representations of the Temperley-Lieb algebra as spin zero
projectors.}

Representations of the TL algebra, commuting with quantum groups, can be
constructed in the following way\cite{R}. Suppose ${\cal U}_q(X_n)$ is the
universal envelopping algebra of a finite dimensional Lie algebra $X_n$,
equipped with the coproduct $\Delta :{\cal U}_q\rightarrow {\cal U}_q\otimes
{\cal U}_q$ \cite{DJ}. If now $\pi :{\cal U}_q\rightarrow EndV_\Lambda $ is
a finite dimensional irreducible representation with highest weight ${%
\Lambda }$ and we assume that the decomposition $V_\Lambda \otimes V_\Lambda
$ is multiplicity free and includes one trivial representation on $V_0$,
then the projector ${\cal P}_0$ from $V_\Lambda \otimes V_\Lambda $ onto $%
V_0 $ is a representation of the TL algebra. The deformation parameter $q$,
which plays the role of a coupling constant in the Hamiltonian, is related
to $Q$ as:
\begin{equation}
\sqrt{Q}=Tr_V(q^{-2\rho }),
\end{equation}
where $\rho $ is half the sum of the positive roots.

By construction ${\cal P}_0$ commutes with the quantum group ${\cal U}%
_q(X_n) $.

Since we are not going to use any group-theoretical machinery, we will just
lift the relevant formulas off Batchelor and Kuniba\cite{BK} in order to
display explicitly  the Hamiltonians to be diagonalized.

We will consider the following specific cases, $(V_\Lambda ,{\cal U}%
_q(X_n))=(V_{2s\Lambda _1},{\cal U}_q(A_1))$ for spin $s$, $(V_{\Lambda _1},%
{\cal U}_q(B_n)$,$(V_{\Lambda _1},{\cal U}_q(C_n)$ and $(V_{\Lambda _1},%
{\cal U}_q(D_n)$. I.e, we treat the $q$-deformations of the spin-$s$
representation of $sl(2)$ and the vector representations of $so(2n+1),sp(2n)$
and $so(2n)$. $V_\Lambda $ denotes the ${\cal U}_q(X_n)$ module with highest
weight $\Lambda $. $\Lambda _1$ is a highest weight of $X_n$.

Introduce the following notation. Let $e_i,i=1,\ldots ,n$ be orthonomal
vectors and express the fundamental weight, $\Pi =\Lambda _1+\ldots +\Lambda
_n$, the set ${\cal A}$ of weights and the coupling constant $\sqrt{Q}\equiv
-2\Delta $ as:

\begin{eqnarray}
A_1:{\cal A}= &&\{s(e_1-e_2),(s-1)(e_1-e_2),\ldots ,-s(e_1-e_2)\},  \nonumber
\\
&&\Lambda _1=(e_1-e_2)/2,  \nonumber \\
&&\rho =(e_1-e_2)/2,  \nonumber \\
&&J=\{s,s-1,\ldots ,-s\},  \nonumber \\
&&\epsilon (\mu )=(-1)^{\tilde \mu },  \nonumber \\
&&\sqrt{Q}=[2s+1];  \nonumber \\
B_n\;(n\geq 2): &{\cal A}=&\{0,\pm e_1,\ldots ,\pm e_n\},  \nonumber \\
&&\Lambda _i=e_1+\ldots +e_i,\;\;\;\;(1\leq i<n),  \nonumber \\
&=&(e_1+\ldots +e_n)/2,\;\;\;\;(i=n),  \nonumber \\
&&\rho =(n-1/2)e_1+\ldots +e_n/2,  \nonumber \\
&&J=\{0,\pm 1,\ldots ,\pm n\},  \nonumber \\
&&\epsilon (\mu )=(-1)^{\tilde \mu },  \nonumber \\
&&\sqrt{Q}=[2n-1][n+1/2]/[n-1/2];  \nonumber \\
C_n:{\cal A}= &&\{\pm e_1,\ldots ,\pm e_n\}, \\
&&\Lambda _i=e_1+\ldots +e_i,\;\;\;\;(1\leq i\leq n),  \nonumber \\
&&\rho =ne_1+\ldots +e_n,  \nonumber \\
&&J=\{\pm 1,\ldots ,\pm n\},  \nonumber \\
&&\epsilon (\mu )=sign(\mu ),  \nonumber \\
&&\sqrt{Q}=[n][2n+2]/[n+1];  \nonumber \\
D_n:{\cal A}= &&\{\pm e_1,\ldots ,\pm e_n\},  \nonumber \\
&&\Lambda _i=e_1+\ldots +e_i,\;\;\;\;(1\leq i<n-1),  \nonumber \\
&=&(e_1+\ldots +e_{n-1}-e_n)/2,\;\;\;\;(i=n-1),  \nonumber \\
&=&(e_1+\ldots +e_{n-1}+e_n)/2,\;\;\;\;(i=n),  \nonumber \\
&&\rho =(n-1)e_1+\ldots +e_{n-1},  \nonumber \\
&&J=\{0,\pm 1,\ldots ,\pm n\},  \nonumber \\
&&\sqrt{Q}=[2n-2][n]/[n-1];  \nonumber \\
&&\epsilon (\mu )=1.  \nonumber
\end{eqnarray}
For $\mu \in J$ the symbol $\tilde \mu $ is defined as $\tilde \mu =\mu
+(1\pm 1)/4$ for $A_1$ with $s\in {\bf Z}+(1\pm 1)/4$ and $\tilde \mu =0$
with the exception of $\tilde 0=1$ for $B_n$. The $q$-number notation is $%
[x]\equiv (q^x-q^{-x})/(q-q^{-1})$ . For $X_n=B_n,C_n,D_n$, we extend the
suffix of $e_\mu $ to $-n\leq \mu \leq n$ by setting $e_{-\mu }=-e_\mu $
(hence $e_0=0$). Using the index set $J$, above. we can write ${\cal A}%
=\{\mu (e_1-e_2)\}$ for $A_1$ and ${\cal A}=\{e_\mu |\mu \in J)\}$ for $%
B_n,C_n,D_n$.

Denoting by $E_{\mu \nu }\in End\,V_\Lambda $ the matrix unit, having all
elements zero, except at row $\mu $ and column $\nu $, the projector can be
written as
\begin{equation}
{\cal P}_0=Q^{-1/2}\sum_{\mu ,\nu \in J}\epsilon (\mu )\epsilon (\nu
)q^{-<e_\mu +e_\nu ,\rho >}E_{\mu \nu }\otimes E_{-\mu -\nu }.
\end{equation}

In the following we will refer to all models generically as {\em higher spin}
models for simplicity, even when not talking about $A_1$

Consider a one-dimensional chain of lenght $N$ with a "spin" at each site.
The spin variables range over the set of weight vectors $v_\mu|\mu\in J$ and
our Hilbert space is an $N$-fold tensor product $V_\Lambda\otimes\ldots%
\otimes V_\Lambda$. For $A_1$, these are the $q$-analogs of the usual spin
states.

The Hamiltonian densities acting on two neighboring sites are then given by:
\begin{equation}
\langle k,l| U | i,j \rangle= \epsilon(i)\epsilon(k)q^{-<e_i+e_k,\rho>}
\delta_{i+j,0}\, \delta_{k+l,0}.
\end{equation}


\section{The coordinate Bethe - Ansatz}

All the above Hamiltonians are $U(1)$ invariant and we can classify their
spectra according to sectors. For $A_1(s=1)$ the commuting operator is the
total spin $S^z=\sum_{k=1}^N S_k^z$ and we set the conserved quantum number $%
r=N-S^z$. In general it equals $r=N*\omega -S^z$ for $A_1$ and $B_n$ and $%
r=N*(\omega -1/2)-S^z$ for $C_n$ and $D_n$. We set $\omega =maxJ$.

Therefore, there exists a reference state $|\Omega\rangle$, satisfying $%
H\left |\Omega \right\rangle = E_0 \left |\Omega \right\rangle$, with $E_0=0$%
. We take $|\Omega\rangle $ to be $|\Omega\rangle = \prod_k^N
\left
|\omega,k \right\rangle$.

In every sector $r$ there are eigenstates degenerate with $\left |\Omega
\right\rangle$. They contain a set of {\em impurities}. We call impurity any
state obtained by {\em lowering} some of the $\left |\omega,k \right\rangle$%
's, such that the sum of any two neighboring spins is non-zero. Since $H(q)$
is a projector on spin zero, all these states are annihilated by $H(q)$. In
particular, they do not {\em move} under the action of $H(q)$, which is the
reason for their name.

We will now start to diagonalize $H(q)$ in every sector. Nothing interesting
happens in sector $r<2\omega $. Sector $r=2\omega $ is more interesting,
although still trivial, since it contains one free pseudoparticle. The main
result of this paper is to show that $H(q)$ can be diagonalized in a
convenient basis, constructed from products of single pseudoparticle
wavefunctions. The energy eigenvalues will be parametrized as a sum of
single pseudoparticle contributions.

\subsection{ The sector $r=2\omega $, containing one pseudoparticle}

Starting with $r=2\omega $, we encounter the situation, where the states $%
\left| j,k\right\rangle $ and $\left| -j,k\pm 1\right\rangle $, $j\neq
\omega $ occur in neighboring pairs. They do move under the action of $H(q)$
and mix with states containing one $\left| -\omega ,k\right\rangle $.
Eigenstates are a superposition of $|x{\scriptstyle[-\omega ]}\rangle
=(\,\ldots \omega \omega \omega \omega \;%
\raisebox{-0.5em}{$\stackrel{\textstyle
-\omega}{\scriptstyle x}$}\,\omega \omega \ldots )$ and $|x{\scriptstyle%
[j,-j]}\rangle =(\ldots \omega \omega \omega \;\raisebox{-0.5em}{$\stackrel{%
\textstyle +j}{\scriptstyle x}$}\,-j\omega \omega \ldots )$, i.e.
\begin{equation}
\left| 2\omega ;\ldots \right\rangle =\sum_x\{a_\omega (x)\left| x{%
\scriptstyle[-\omega ]}\right\rangle \,+\,\sum_j^{\prime }b_j(x)\left| x{%
\scriptstyle[j,-j]}\right\rangle \},
\end{equation}
where $\sum_j^{\prime }$ means $j\in J^{*}=J-\{\pm \omega \}$ and the
ellipses stand for parameters the eigenvector is going to depend on. When $%
H(q)$ now acts on $\left| 2\omega ;\ldots \right\rangle $ it sees the
reference configuration, except in the vicinity of $x$ and we obtain the
eigenvalue equations

\begin{eqnarray}
&&(E-\qplus -\qmins )\;a_\omega (x)=a_\omega (x+1)\,+\,a_\omega (x-1)\,+
\nonumber \\
&&\;\;\;\;\,\sum_l^{\prime }\epsilon (\omega )\epsilon (l)q^{-<e_\omega
+e_l,\rho >}\,b_l(x-1)\,+\,\sum_l^{\prime }\epsilon (-\omega )\epsilon
(l)q^{-<e_{-\omega }+e_l,\rho >}\,\,b_l(x)  \nonumber \\
&&  \nonumber \\
&&Eb_j(x)=\epsilon (\omega )\epsilon (l)q^{-<e_j+e_\omega ,\rho >}\,a_\omega
(x+1)\,+\,\epsilon (-\omega )\epsilon (l)q^{-<e_j-e_{-\omega },\rho
>}\,\,a_\omega (x)+  \nonumber \\
\;\;\;\; &&\sum_l^{\prime }\epsilon (j)\epsilon (l)q^{-<e_j+e_l,\rho
>}b_l(x),\;\;\;\,j\in J^{*}.  \label{eq:f}
\end{eqnarray}
Eliminating the $b_j$'s, we get an equation very similar to the XXZ model:
\begin{equation}
(E-\sum_{j\in J}q^{-2<e_l,\rho >})a_\omega (x)=a_\omega (x+1)+a_\omega (x-1).
\end{equation}

We will treat periodic boundary conditions maintaining translational
invariance in the following sections. They demand $a_\omega
(x+N)\,=\,a_\omega (x)$ and $b_j(x+N)\,=\,b_j(x)$. We parametrize as: $%
a_\omega (x)=a_\omega \xi ^x$ and $b_l(x)=b_l\xi ^x,\;\;l\in J^{*}$.
Substituting this into equ.(\ref{eq:f}) we get two eigenstates and their
energies
\begin{eqnarray}
a_w &=&\epsilon (-\omega )q^{<e_\omega ,\rho >}+\epsilon (\omega
)q^{-<e_\omega ,\rho >}\xi ^{-1}\equiv \Gamma (\xi ^{-1})  \label{r21} \\
b_l &=&\epsilon (l)q^{-<e_l,,\rho >},\,l\in J^{*} \\
\smallskip E_1 &=&\sum_l^{\prime }q^{-<e_l,\rho >}+\Gamma (\xi )\Gamma ({\xi
}^{-1})=\sum_{l\in J}q^{-2<e_l,\rho >}+\xi +\xi ^{-1}  \nonumber
\end{eqnarray}
and a highly degenerate solution with $E_2=0$, with the following constraint
on the parameters:
\begin{equation}
\sum_l^{\prime }\epsilon (l)\epsilon (\omega )q^{-<e_{l+\omega },\rho
>}b_l+\Gamma (\xi )\epsilon (\omega )q^{-<e_\omega ,\rho >}a_\omega =0.
\end{equation}
Here ${\xi }=e^{i\theta }$, $\theta $ being the momentum determined from the
periodic condition to be: $\theta =2{\pi }l/N$, with $l$ integer.

We describe this situation by saying that we have two types of
pseudoparticles with energies $E_1$ and $E_2$. Whereas the pseudoparticle $%
\left| 2\omega ;\theta \right\rangle _2$ is degenerate with $|\Omega \rangle
$, i.e. propagates with energy $E_2=0$, the pseudoparticle $|2\omega ;\theta
\rangle _1$, propagates with energy
\begin{equation}
E_1\;=-2\Delta +2\cos \theta ,\;\;\;\;\,2\Delta \equiv -\sum_{l\in
J}q^{-2<e_l,\rho >}.  \label{paren}
\end{equation}
As mentioned before, the energy eigenvalues are going to be parametrized as
a sum of single pseudoparticle energies. Thus we write:
\begin{equation}
E=\sum_{n=1}^p\epsilon _n(\sum_l^{\prime }q^{-<e_l,\rho >}+\Gamma (\xi
_n)\Gamma ({\xi }_n^{-1})),  \label{energy}
\end{equation}
where $\epsilon _n$ depends on which pseudoparticle we use: $\epsilon _n=1$
for $E=E_1$ and $E=E_2=0$.


\subsection{ Two pseudoparticles and the XXZ Bethe - Ansatz}

The next higher sector would be $r=2\omega +1$, but let us treat $r=4\omega $
first, since then we can compare it with the first nontrivial sector in the
XXZ model.

This sector contains states, which consist of two interacting
pseudoparticles. We seek these eigenstates in the form:
\begin{eqnarray}
&&\left| 4\omega ;\ldots \right\rangle _{\epsilon _1\epsilon
_2}=\sum_{x_1<x_2}\{a_{\omega \omega }(x_1,x_2)\left| x_1{\scriptstyle %
[-\omega ]},x_2{\scriptstyle [-\omega ]}\right\rangle +\sum_i^{\prime
}b_{\omega i}(x_1,x_2)\left| x_1{\scriptstyle [-\omega ]},x_2{\scriptstyle %
[i,-i]}\right\rangle +  \nonumber  \label{4param} \\
&&\sum_j^{\prime }b_{j\omega }(x_1,x_2)\left| x_1{\scriptstyle [j,-j]},x_2{%
\scriptstyle [-\omega ]}\right\rangle +\sum_i^{\prime }\sum_j^{\prime
}b_{ij}(x_1,x_2)\left| x_1{\scriptstyle [i,-i]},x_2{\scriptstyle [j,-j]}%
\right\rangle \}.\
\end{eqnarray}
Translational invariance now specifies $a_{\omega \omega }(x_1,x_2)={\xi }%
^{x_1}a_{\omega \omega }(n)$ and similarly for the other wave functions,
where $n=x_2-x_1$. Periodic boundary conditions require that
\begin{eqnarray}
a_{\omega \omega }(n) &=&{\xi }^na_{\omega \omega }(N-n),\hspace{1.5cm}
\nonumber \\
b_{i\omega }(n) &=&{\xi }^nb_{\omega i}(N-n),\;b_{ij}(n)={\xi }^nb_{ji}(N-n),
\label{eq:3k}
\end{eqnarray}
where ${\xi }={\xi }_1{\xi }_2$ \thinspace (${\xi }_i=e^{i\theta _i}$, $%
i=1,2 $) and the total momentum is $\theta _1+\theta _2=2{\pi }l/N$, with $l$
integer.

According to equ.(\ref{energy}), we will parametrize the energy as
\begin{equation}
E=\sum_{n=1}^2\epsilon _n[\sum_l^{\prime }q^{-<e_l,\rho >}+\Gamma (\xi
_n)\Gamma ({\xi }_n^{-1})].  \label{eig4}
\end{equation}

Let us take the block $\epsilon_1=\epsilon_2=1$ first. We try to build $2$%
-pseudoparticle eigenstates out of translationally invariant products of $1$%
-pseudoparticle excitations at $x_1$ and $x_2$ with weight functions $%
D_{i}(x_1,x_2)$, $i=1,2$:

\begin{eqnarray}
&&\left| 4\omega ;\theta _1,\theta _2\right\rangle
_{11}=\sum_{x_1<x_2}\{\;D_1(x_1,x_2)[\Gamma (\xi _1^{-1})|x_1{\scriptstyle %
[-\omega ]}\rangle +\sum_i^{\prime }\epsilon (i)q^{-<e_i,\rho >}|x_1{%
\scriptstyle [i,-i]}\rangle \,]  \nonumber \\
&&\lbrack \;\Gamma (\xi _2^{-1})|x_2{\scriptstyle [-\omega ]}\rangle
+\sum_j^{\prime }\epsilon (j)q^{-<e_j,\rho >}|x_2{\scriptstyle [j,-j]}%
\rangle \,]+  \nonumber \\
&&\;D_2(x_1,x_2)[\,\Gamma (\xi _2^{-1})|x_1{\scriptstyle [-\omega ]}\rangle
+\sum_j^{\prime }\epsilon (j)q^{-<e_j,\rho >}|x_1{\scriptstyle [j,-j]}%
\rangle \,]  \nonumber \\
&&\lbrack \,\Gamma (\xi _1^{-1})|x_2{\scriptstyle [-\omega ]}\rangle
+\sum_i^{\prime }\epsilon (i)q^{-<e_i,\rho >}|x_2{\scriptstyle [i,-i]}%
\rangle \,]\;\;\}.  \nonumber
\end{eqnarray}

Comparing this with equ.(\ref{4param}) and using translational invariance,
implying $D_2(n)=\xi^n D_1(N-n)$, we get
\begin{eqnarray}  \label{4reps}
a_{\omega\omega}(n)&=&\Gamma(\xi^{-1}_{1})\,\Gamma(\xi^{-1}_{2})\,D(n),
\nonumber \\
b_{\omega i}(n)&=&\epsilon(i)\, q^{-<e_i,\rho>}
[\,\Gamma(\xi^{-1}_{1})\,D_{1}(n)+ \Gamma(\xi^{-1}_{2})\,D_{2}(n)],
\nonumber \\
\ b_{i\omega }(n)&=&\epsilon(i) q^{-<e_i,\rho>}
[\,\Gamma(\xi^{-1}_{2})D_{1}(n)+ \Gamma(\xi^{-1}_{1}\,)D_{2}(n)\,],
\nonumber \\
\ b_{ ij}(n)&=&\epsilon(i)\epsilon(j)\, q^{-<e_i+e_j,\rho>}\, D(n),\;\;\; 3
\leq n\leq N-3,
\end{eqnarray}
where $D(n)=D_{1}(n)+D_{2}(n)$.

Applying $H(q)$ to the state of (\ref{4param}), we obtain a set of coupled
equations for $a_{\omega \omega }(n)$,\thinspace $b_{ij}(n)$. Following \cite
{Park}, we split the equations into {\em far} equations, when excitations do
not meet and {\em near} equations, containing terms when they are neighbors.
The far equations are:
\begin{eqnarray}
&&(E-2q^{-2<e_\omega ,\rho >}-2q^{2<e_\omega ,\rho >})a_{\omega \omega
}(n)=(1+\xi )^{-1})a_{\omega \omega }(n+1)+(1+\xi )a_{\omega \omega }(n-1)+
\nonumber  \label{4fareq} \\
&&\mbox{}\sum_l^{\prime }\epsilon (l)\epsilon (\omega )q^{-<e_l+e_\omega
,\rho >}[\,\xi ^{-1}b_{l\omega }(n+1)+b_{\omega l}(n-1)\,]+  \nonumber \\
&&\sum_l^{\prime }\epsilon (l)\epsilon (-\omega )q^{-<e_l-e_\omega ,\rho
>}[\,b_{l\omega }(n)+b_{\omega l}(n)\,],\;\;\;2\leq n\leq N-2,\\
\vspace{.3cm}
&&(E-q^{-2<e_\omega ,\rho >}-q^{<2e_\omega ,\rho >})b_{\omega j}(n)={\xi }%
^{-1}b_{\omega j}(n+1)+{\xi }b_{\omega j}(n-1)+  \nonumber \\
&&\mbox{}\epsilon (j)\epsilon (-\omega )q^{-<e_j-e_\omega ,\rho
>}\,a_{\omega \omega }(n)+\epsilon (j)\epsilon (\omega )q^{-<e_j+e_\omega
,\rho >}\,a_{\omega \omega }(n+1)+  \nonumber \\
&&\sum_l^{\prime }\epsilon (l)q^{-<e_l,\rho >}\left[ \epsilon (\omega
)q^{-<e_\omega ,\rho >}\xi ^{-1}b_{lj}(n+1)+\epsilon (-\omega )q^{<e_\omega
,\rho >}b_{lj}(n)+\right.  \nonumber \\
&&\left. \epsilon (j)q^{-<e_j,\rho >}\xi ^{-1}b_{\omega l}(n)\right]
,\;\;\;2\leq n\leq N-2,\\
\vspace{.3cm}
&&(E-q^{-2<e_\omega ,\rho >}-q^{2<e_\omega ,\rho >})b_{j\omega
}(n)=b_{j\omega }(n-1)+b_{\omega j}(n+1)+  \nonumber \\
&&\mbox{}\epsilon (j)\epsilon (-\omega )q^{-<e_j-e_\omega ,\rho
>}\,a_{\omega \omega }(n)+\epsilon (j)\epsilon (\omega )q^{-<e_j+e_\omega
,\rho >}\xi \,a_{\omega \omega }(n-1)+  \nonumber \\
&&\sum_l^{\prime }\epsilon (l)q^{-<e_l,\rho >}\left[ \epsilon (\omega
)q^{-<e_\omega ,\rho >}b_{jl}(n-1)+\epsilon (-\omega )q^{<e_\omega ,\rho
>}b_{jl}(n)+\epsilon (j)q^{-<e_j,\rho >}b_{l\omega }(n)\right] ,  \nonumber
\\
&&\;\;\;3\leq n\leq N-3,\\
\vspace{.3cm}
&&Eb_{ij}(n)=\epsilon (\omega )q^{-<e_\omega ,\rho >}[\epsilon
(i)q^{-<e_i,\rho >}\xi b_{\omega j}(n-1)+\epsilon (j)q^{-<e_j,\rho
>}b_{\omega i}(n+1)]+  \nonumber \\
&&\mbox{}\epsilon (-\omega )q^{<e_\omega ,\rho >}[\epsilon (i)q^{-<e_i,\rho
>}b_{\omega j}(n)+\epsilon (j)q^{-<e_j,\rho >}b_{\omega i}(n)]+  \nonumber \\
&&\sum_l^{\prime }\epsilon (l)q^{-<e_l,\rho >}[\epsilon (i)q^{-<e_i,\rho
>}b_{lj}(n)+\epsilon (j)q^{-<e_j,\rho >}b_{il}(n)],  \nonumber \\
&&\mbox{}3\leq n\leq N-3.
\end{eqnarray}

We already know them to be satisfied, if we parametrize $D_1(n)$ and $D_2(n)$
by plane waves:
\begin{equation}
D_1(n)\,=\,\xi_{2}^{n},\;D_2(n)=\xi_{2}^{N}\xi_{1}^{n}.  \label{eq:4ll}
\end{equation}

The real problem arises of course, when pseudoparticles are neighbors, so
that they interact and we have no guarantee that the total energy is a sum
of single pseudoparticle energies.


Let us now have a lightning review of the XXZ coordinate BA in order to be
able to comment on the features, which are not going to survive
generalizations to the present models.

The equations in the sector $r_{XXZ}=2$ are
\begin{equation}
(E-2q^{\frac 12}-2q^{-\frac 12})\;a(x_1,x_2)=  \nonumber  \label{fXXZ}
\end{equation}
\vspace{-.4cm}
\begin{equation}
a(x_1+1,x_2)+a(x_1-1,x_2)+a(x_1,x_2+1)+a(x_1,x_2-1),  \nonumber
\end{equation}
if $x_1$ and $x_2$ are not neighbors. In case they are we get
\begin{equation}
(E-q^{\frac 12}-q^{-\frac 12})\;a(x_1,x_1+1)=a(x_1-1,x_1+1)+a(x_1,x_1+2).
\label{nXXZ}
\end{equation}
One now supposes, that the parametrization equ.(\ref{4param}) for $s=1/2$
solves {\em both} the above equations. In this case we are allowed to set $%
x_2=x_1+1$ in equ.(\ref{fXXZ}) and subtract it from equ.(\ref{nXXZ}),
yielding the following consistency condition :
\begin{equation}
-(q^{\frac 12}+q^{-\frac 12})a(x_1,x_1+1)=a(x_1,x_1)+a(x_1+1,x_1+1).
\end{equation}

This gives the {\em BA equation} for the XXZ model, determining the 2-body
phase shift:
\begin{equation}
\xi_2^N = - \frac{ 1+\xi+\xi_2(q^{\frac{1}{2}} +q^{- \frac{1}{2}} ) } {
1+\xi+\xi_1(q^{\frac{1}{2}} +q^{-\frac{1}{2}} ) }.
\end{equation}

This type of procedure only works for the spin 1/2 XXZ model, due to the
following fact. When the two pseudoparticles come together in a
configuration like $(\ldots+++--+++\ldots)$ and when H is applied to the two
down spins, it gives zero, since their total $S_z$ equals $-1$. But whenever
two excitations approach each other becoming neighbors and the Hamiltonian
applied to them yields a nonvanishing resul t, then the representation like
equ.(\ref{4reps}) cannot solve both the {\em far} and {\em near} equations.
Yet this is exactly the situation arising for higher spins. As we shall see,
in this case, the representation equ.(\ref{4reps}) has to modified\cite{KL},
the two-body wavefunction developing a "discontinuity" at minimum
separation. We call this the {\em spin zero rule}.

Now back to our problem to solve the near equations. They are:
\begin{eqnarray}
&&(E-q^{-2<e_\omega ,\rho >}-q^{2<e_\omega ,\rho >})a_{\omega \omega
}(1)=(1+\xi )^{-1}a_{\omega \omega }(2)+  \nonumber  \label{4neareq} \\
&&\;\;\;\sum_l^{\prime }\epsilon (l)q^{-<e_l,\rho >}[\epsilon (\omega
)q^{-<e_\omega ,\rho >}\,\xi ^{-1}b_{l\omega }(2)+\epsilon (-\omega
)q^{<e_\omega ,\rho >}\,b_{\omega l}(1)\,];  \nonumber
\end{eqnarray}
\begin{eqnarray}
&&(E-q^{-2<e_\omega ,\rho >})b_{\omega j}(1)={\xi }^{-1}b_{\omega j}(2)+
\nonumber \\
&&\;\;\;\epsilon (j)\epsilon (-\omega )q^{-<e_j-e_\omega ,\rho >}\,a_{\omega
\omega }(1)+\epsilon (j)\epsilon (\omega )q^{-<e_j+e_\omega ,\rho
>}\,a_{\omega \omega }(2)+  \nonumber \\
&&\sum_l^{\prime }\epsilon (l)q^{-<e_l,\rho >}\left[ \epsilon (\omega
)q^{-<e_\omega ,\rho >}\xi ^{-1}b_{lj}(2)+\epsilon (j)q^{-<e_j,\rho
>}b_{\omega l}(1)\right] ;  \nonumber
\end{eqnarray}
%
\begin{eqnarray}
&&(E-q^{2<e_\omega ,\rho >})b_{j\omega }(2)=b_{\omega j}(3)+  \nonumber \\
&&\;\;\;\epsilon (j)\epsilon (-\omega )q^{-<e_j-e_\omega ,\rho >}\,a_{\omega
\omega }(2)+\epsilon (j)\epsilon (\omega )q^{-<e_j+e_\omega ,\rho >}\xi
\,a_{\omega \omega }(1)+  \nonumber \\
&&\sum_l^{\prime }\epsilon (l)q^{-<e_l,\rho >}\left[ \epsilon (-\omega
)q^{<e_\omega ,\rho >}b_{jl}(2)+\epsilon (j)q^{-<e_j,\rho >}b_{l\omega
}(2)\right] ;  \nonumber
\end{eqnarray}
\begin{eqnarray}
&&Eb_{ij}(2)=\epsilon (\omega )q^{-<e_\omega ,\rho >}[\epsilon
(i)q^{-<e_i,\rho >}\xi b_{\omega j}(1)+\epsilon (j)q^{-<e_j,\rho
>}b_{i\omega }(3)]+  \nonumber  \label{bij} \\
&&\;\;\;\epsilon (-\omega )q^{<e_\omega ,\rho >}[\epsilon (i)q^{-<e_i,\rho
>}b_{\omega j}(2)+\epsilon (j)q^{-<e_j,\rho >}b_{i\omega }(2)]+  \nonumber \\
&&\sum_l^{\prime }\epsilon (l)q^{-<e_l,\rho >}[\epsilon (i)q^{-<e_i,\rho
>}b_{lj}(2)+\epsilon (j)q^{-<e_j,\rho >}b_{il}(2)],  \nonumber \\
&&\mbox{}i\neq j;  \nonumber
\end{eqnarray}
\begin{eqnarray}
&&Eb_{ii}(2)=\epsilon (i)\epsilon (\omega )q^{-<e_i+e_\omega ,\rho >}[\xi
b_{\omega i}(1)+b_{i\omega }(3)]+  \nonumber  \label{bi} \\
&&\epsilon (i)\epsilon (-\omega )q^{-<e_i-e_\omega ,\rho >}[b_{\omega
i}(2)+b_{i\omega }(2)]+\sum_l^{\prime }\epsilon (l)q^{-<e_i+e_l,\rho
>}[b_{li}(2)+b_{il}(2)]+  \nonumber \\
&&\sum_l^{\prime }\epsilon (-i)\epsilon (l)q^{-<e_l-e_i,\rho >}{\cal B}%
_i^{(l)}.
\end{eqnarray}

Here some new states are showing up. ${\cal B}_i^{(l)}$ are the
wavefunctions of the states of the type $(\ldots \;\omega \omega
\raisebox{-0.5em}{$\stackrel{\textstyle i}
{\scriptstyle x}$}\;l\;-l\;\omega \omega \ldots ),l\neq i$. Applying $H(q)$
to them we obtain the system:
\begin{equation}
(E-q^{2<e_l,\rho >}){\cal B}_i^{(l)}=\sum_{j\in J^{*},j\neq -i}\epsilon
(l)\epsilon (j)q^{_{<}e_l+e_j,\rho >}{\cal B}_i^{(l)}+\epsilon (l)\epsilon
(-i)q^{-<e_l-e_i,\rho >}b_{ii}(2),  \nonumber
\end{equation}
yielding
\begin{equation}
{\cal B}_i^{(l)}=\frac{\epsilon (l)\epsilon (-i)q^{<e_l-e_i,\rho >}}{%
(E-\sum_{j\in J,j\neq -i}q^{-2<e_j,\rho >})}b_{ii}(2).
\end{equation}
Eliminating ${\cal B}_i^{(l)}$ from equation (\ref{bi}), we get
\begin{eqnarray}
&&\frac{E(E+2\Delta )}{E+2\Delta +q^{<e_i,\rho >}}\;b_{ii}(2)=\epsilon
(i)\epsilon (\omega )q^{-<e_i+e_\omega ,\rho >}[\xi b_{\omega
i}(1)+b_{i\omega }(3)]+  \nonumber  \label{bii} \\
&&\epsilon (i)\epsilon (-\omega )q^{-<e_i-e_\omega ,\rho >}[b_{\omega
i}(2)+b_{i\omega }(2)]+  \nonumber \\
&&\sum_l^{\prime }\epsilon (l)\epsilon (i)q^{-<e_i+e_l,\rho
>}[b_{li}(2)+b_{il}(2)].
\end{eqnarray}
In order to solve these equations, we follow \cite{KL} and now leave the
value of the wavefunctions for nearest separation as arbitrary parameters:
%
%
\begin{eqnarray}
a_{\omega \omega }(1) &=&\Gamma (\xi )\Gamma (\xi ^{-1})D(1)+{\cal F}%
_{a_{\omega \omega }}(1),  \nonumber  \label{4Fequs} \\
b_{\omega i}(1) &=&\epsilon (i)q^{-<e_i,\rho >}[\Gamma (\xi
^{-1})D_1(1)+\Gamma (\xi ^{-1})D_2(1)]+{\cal F}_{b_{\omega i}}(1),  \nonumber
\\
b_{\omega i}(2) &=&\epsilon (i)q^{-<e_i,\rho >}[\Gamma (\xi
^{-1})D_1(2)+\Gamma (\xi ^{-1})D_2(2)]+{\cal F}_{b_{i\omega }}(2),  \nonumber
\\
b_{ij}(2) &=&\epsilon (i)\epsilon (j)q^{-<e_i+e_j\rho >}D(2)+{\cal F}%
_{b_{ij}}(2).
\end{eqnarray}

In order for this modification to leave the far equations still satisfied,
the following conditions have to hold:
\begin{eqnarray}
&&(1+\xi ^{-1}){\cal F}_{a_{\omega \omega }}(1)+\sum_l^{\prime }\epsilon
(l)\epsilon (\omega )q^{-<e_l+e_\omega ,\rho >}{\cal F}_{b_{\omega l}}(1)+
\nonumber  \label{Fcons} \\
&&\;\;\;\;\sum_l^{\prime }\epsilon (l)\epsilon (-\omega )q^{-<e_l-e_\omega
,\rho >}{\cal F}_{b_{l\omega }}(2)=0,  \nonumber \\
&&\xi {\cal F}_{b_{\omega j}}(1)+\sum_l^{\prime }\epsilon (l)\epsilon
(-\omega )q^{-<e_l-e_\omega ,\rho >}{\cal F}_{b_{lj}}(2)=0,  \nonumber \\
&&{\cal F}_{b_{j\omega }}(2)+\sum_l^{\prime }\epsilon (l)\epsilon (\omega
)q^{-<e_l+e_\omega ,\rho >},{\cal F}_{b_{jl}}(2)=0.
\end{eqnarray}
Now using equs.(\ref{4reps}) and equ.(\ref{4Fequs}) in equ.(\ref{bii}) , we
get the following equation for ${\cal F}_{b_{ii}}(2)$:
\begin{equation}
{\cal F}_{b_{ii}}(2)=\frac{D(2)}{E+2\Delta },\;\;i\in J^{*}.  \label{Fbii}
\end{equation}
Doing the same with equ.(\ref{bij}), we get:
\begin{equation}
{\cal F}_{b_{ij}}(2)=0,\;\;\;i\neq j.  \label{Fbij}
\end{equation}
These results for ${\cal F}_{b_{ii}}(2),{\cal F}_{b_{ij}}(2)$ are
reasonable. In the first case the colliding excitations satisfy the {\em %
zero spin rule} and we get a non-zero result, wheras in the second case the
rule is not satisfied and we get zero. Using this in equ.(\ref{Fcons}), we
obtain for the remaining constants:
\begin{eqnarray}
{\cal F}_{a_{\omega \omega }}(1) &=&-(q^{-2<e_\omega ,\rho >}+q^{2<e_\omega
,\rho >}+2\Delta )\;{\cal F}_{b_{ii}}(2),  \nonumber  \label{Fabb} \\
{\cal F}_{b_{\omega j}}(1) &=&-\epsilon (j)\epsilon (-\omega
)q^{-<e_j-e_\omega ,\rho >}\;{\cal F}_{b_{ii}}(2),  \nonumber \\
{\cal F}_{b_{j\omega }}(2) &=&-\epsilon (j)\epsilon (\omega
)q^{-<e_j+e_\omega ,\rho >}\;{\cal F}_{b_{ii}}(2)\;,\;j\in J*.
\end{eqnarray}
Substituting finally the complete parametrization into the remaining near
equatins, get the following Bethe-Ansatz equation:
\begin{equation}
\frac{D(2)}{E+2\Delta }=\frac \xi {1+\xi }D(1),
\end{equation}
which can also be rewritten as
\begin{equation}
\xi _2^N=-\frac{\xi _2[(1+\xi ^{-1})\xi _2-2\Delta -E]}{\xi _1[(1+\xi
^{-1})\xi _1-2\Delta -E]}.
\end{equation}
Using the explicit form of the energy, the set of equations determining the
spectrum are:
\begin{equation}
\xi _2^N=-\frac{1+\xi \xi _2-2\Delta \xi _2}{1+\xi \xi _1-2\Delta \xi _1}%
,\;\;\;\xi ^N=1.
\end{equation}

Notice that this equation is independent of $n$ or any other representation
specific quantities. All the models considered show therefore an equivalent
spectrum, when parametrized in terms of $\Delta$.

In particular, this is the same consistency condition one finds for the XXZ
model, showing that for $\epsilon_1=\epsilon_2=1$, even for periodic
boundary conditions, the spectra of all our models are equivalent to the
spectrum of the XXZ model, if expressed in terms of $\Delta$ .

We will refrain from discussing the other two blocks: $\epsilon
_1=1,\epsilon _2=0$ and $\epsilon _1=\epsilon _2=0$, since the calculations
are analogous to the ones presented above. For details the reader might
consult reference\cite{KL}. Anyhow, the first of the two cases doesn't show
up for the more interesting situation of free boundary conditions. It is too
asymetric to satisfy free boundary conditions. This is the reason, why for
free boundary conditions, the spectrum of all of our models is equivalent to
the XXZ spectrum. The block $\epsilon _1=\epsilon _2=0$ has $E=0$ and the BA
equation reduces to $\xi ^N=1$, being highly degenerate. The eigenvalue $%
E=0$ also occurs in the XXZ spectrum, albeit with different degeneracy.

\subsection{ One pseudoparticle and impurities}

Since the setup with pseudoparticles and impurities is a little different
from the case of two pseudoparticles, we will dedicate some space to it.

The eigenstates sought for, will be like:

\begin{eqnarray}
&&\left| 2\omega +1;\ldots \right\rangle =\sum_{x_1<x_2}\{a_{\omega
k}(x_1,x_2)\left| x_1{\scriptstyle [-\omega ]},x_2{\scriptstyle [k]}%
\right\rangle +\sum_i^{\prime }b_{jk}(x_1,x_2)\left| x_1{\scriptstyle [j,-j]}%
,x_2{\scriptstyle [k]}\right\rangle +  \nonumber  \label{3param} \\
&&a_{k\omega }(x_1,x_2)\left| x_1{\scriptstyle [-k]},x_2{\scriptstyle %
[-\omega ]}\right\rangle +\sum_j^{\prime }b_{kj}(x_1,x_2)\left| x_1{%
\scriptstyle [k]},x_2{\scriptstyle [j,-j]}\right\rangle .
\end{eqnarray}

Translational invariance and periodic boundary conditions impose:
\begin{equation}
a_{\omega k}(x_1,x_2) = {\xi}^{x_1} a_{\omega k}(n),\;\; b_{jk}(x_1,x_2) = {%
\xi}^{x_1} b_{j k}(n),\;\;
\end{equation}
\begin{equation}
a_{\omega k}(n) = {\xi}^n a_{k\omega}(N-n),\hspace{1.5cm} b_{jk}(n)={\xi}%
^nb_{kj}(N-n),
\end{equation}
where $n=x_2-x_1, \;{\xi}={\xi}_{1}{\xi}_{2}$ \,(${\xi}_{i}= e^{i\theta_{i}}$%
, $i=1,2$) and the total momentum is $\theta_1+\theta_2=2{\pi}l/N$, with $l$
integer.

Let us take the block $\epsilon_1=1$ , building eigenstates out of
translationally invariant products of $1$-pseudoparticle excitations at $x_1$
and an impurity at $x_2$ with weight functions $D_{i}(x_1,x_2)$, $i=1,2$ as
in the previous section. This yields the parametrizations:
\begin{eqnarray}
\label{3reps}
a_{\omega k}(n)&=&\Gamma(\xi^{-1}_{1})\,D_1(n),  \nonumber \\
a_{k\omega }(n)&=&\Gamma(\xi^{-1}_{1})\,D_2(n),  \nonumber \\
b_{jk}(n)&=&\epsilon(j)\, q^{-<e_j,\rho>} \,D_{1}(n),  \nonumber \\
b_{kj}(n)&=&\epsilon(j)\, q^{-<e_j,\rho>} \,D_{2}(n).
\label{4eq}
\end{eqnarray}

The far equations for the impurity at the right are now:
\begin{eqnarray}
&&(E-2q^{-2<e_\omega ,\rho >}-2q^{2<e_\omega ,\rho >})a_{\omega
k}(x_1,x_2)=a_{\omega k}(x_1-1,x_2))+a_{\omega k}(x_1+1,x_2)+  \nonumber
\label{3fareq} \\
&&\mbox{}\sum_l^{\prime }\epsilon (l)\epsilon (\omega )q^{-<e_l+e_\omega
,\rho >}b_{lk}(x_1-1,x_2)+\sum_l^{\prime }\epsilon (l)\epsilon (-\omega
)q^{-<e_l-e_\omega ,\rho >}\,b_{lk}(x_1,x_2),  \nonumber \\
&&x_1+2\leq x_2\leq N-x_1-2,\\
&&Eb_{jk}(x_1,x_2)=\epsilon (j)\epsilon (\omega )q^{-<e_j+e_\omega ,\rho
>}a_{\omega k}(x_1+1,x_2)+  \nonumber \\
&&\epsilon (j)\epsilon (-\omega )q^{-<e_j-_\omega ,\rho >}a_{\omega
k}(x_1,x_2)+  \nonumber \\
&&\sum_l^{\prime }\epsilon (j)\epsilon (l)q^{-<e_j+e_l,\rho >}[\epsilon
(i)q^{-<e_i,\rho >}b_{lk}(x_1,x_2),  \nonumber \\
&&\mbox{}x_1+3\leq x_2\leq N-x_1-3,
\end{eqnarray}
and analogous equations for the impurity at the left. Eliminating the $b$%
-functions, we get:
\begin{eqnarray}
&&(E-\sum_{l\in J}q^{-2<e_l,\rho >})a_{\omega k}(x_1,x_2)=a_{\omega
k}(x_1-1,x_2)+a_{\omega k}(x_1+1,x_2),  \nonumber \\
&&(E-\sum_{l\in J}q^{-2<e_l,\rho >})a_{k\omega }(x_1,x_2)=a_{k\omega
}(x_1,x_2-1)+a_{k\omega }(x_1,x_2+1),  \nonumber \\
&&\mbox{}x_1+3\leq x_2\leq N-x_1-3.
\end{eqnarray}

We know them to be satisfied, if the energy is given by equ.({\ref{eig4})}. The
 near equations require of course the by now costumary
treatment of modifying the ansatz of the wavefunctions at nearest
separations.

The near  equations for the impurity at the right are:
\[
(E-2q^{-2<e_\omega ,\rho >})a_{\omega k}(x,x+1)=
\]
\begin{equation}
a_{\omega k}(x-1,x+1))+\mbox{}\sum_l^{\prime }\epsilon (l)\epsilon (\omega
)q^{-<e_l+e_\omega ,\rho >}b_{lk}(x-1,x+1),  \label{3neareq}
\end{equation}
\vspace{-.5cm}
\begin{eqnarray}
&&Eb_{jk}(x,x+1)=\epsilon (j)\epsilon (\omega )q^{-<e_j+e_\omega ,\rho
>}a_{\omega k}(x+1,x+2)+  \nonumber \\
&&\epsilon (j)\epsilon (-\omega )q^{-<e_j-_\omega ,\rho >}a_{\omega
k}(x,x+2)+ \sum_l^{\prime }\epsilon (j)\epsilon (l)q^{-<e_j+e_l,\rho
>}b_{lk}(x,x+2),\\
&&Eb_{jj}(x,x+2)=  \nonumber \\
&&\epsilon (j)\epsilon (\omega )q^{-<e_j+e_\omega ,\rho >}a_{\omega
j}(x+1,x+2)+\epsilon (j)\epsilon (-\omega )q^{-<e_j-_\omega ,\rho
>}a_{\omega j}(x,x+2)+  \nonumber \\
&&\epsilon (-j)\epsilon (-\omega )q^{<e_j+e_\omega ,\rho >}a_{j\omega
}(x,x+1)+\epsilon (-j)\epsilon (\omega )q^{<e_j-_\omega ,\rho >}a_{j\omega
}(x,x+2)+  \nonumber \\
&&\sum_l^{\prime }\epsilon (j)\epsilon (l)q^{-<e_j+e_l,\rho
>}b_{lj}(x,x+2)+\sum_l^{\prime }\epsilon (-j)\epsilon (l)q^{<e_j-e_l,\rho
>}b_{jl}(x,x+1).
\end{eqnarray}

They can be solved modifying the parametrization for nearest neighbors in
the usual way. The result is:
\begin{eqnarray}
a_{\omega k}(1) &=&\Gamma (\xi ^{-1})\xi _2+{\cal F}_{a_{k\omega
}},\;\;b_{jk}(2)=\epsilon (j)q^{-<e_j,\rho >}\xi _2^2+{\cal F}_{b_{jk}}, \\
a_{k\omega }(1) &=&\Gamma (\xi ^{-1})\xi _2^N\xi _1+{\cal F}_{a_{k\omega
}},\;\;b_{kj}(1)=\epsilon (j)q^{-<e_j,\rho >}\xi _2^N\xi _1+{\cal F}_{b_{kj}},
\end{eqnarray}
where
\begin{eqnarray}
{\cal F}_{a_{k\omega }} &=&-\epsilon (\omega )q^{-<e_\omega ,\rho >}\xi _2^N,%
{\cal F}_{b_{kj}}=\epsilon (k)q^{-<e_k,\rho >}\xi _2^N\delta _{k+j,0},
\nonumber \\
{\cal F}_{a_{\omega k}} &=&-\epsilon (-\omega )q^{<e_\omega ,\rho >}\xi _2,%
{\cal F}_{b_{jk}}=\epsilon (-k)q^{<e_k,\rho >}\xi \xi _2\delta _{k-j,0},
\end{eqnarray}
with $\xi ^N=1,\xi _1^{N-2}\xi ^2=1$ resulting from periodic boundary
conditions.

\section{ Free boundary conditions}

It is for free boundary conditions, that the Hamiltonian $H(Q)$ commutes
with the quantum group ${\cal U}_q (X_n)$. As expected, the Bethe states are
highest weight states of ${\cal U}_q (X_n)$, except some $E=0$ states. Since
the extension of the BA-procedure from the periodic boundary conditions to
the free case, follows exactly the lines of ref.\cite{KL}, we will only
state the results for the sector $r=4$.

Take the block $\epsilon _1=\epsilon _2=1$. The nearest approach constants
to be added to the now standing waves are the same as in the periodic case,
namely equs.(\ref{Fbij}),(\ref{Fabb}), only ${\cal F}_{b_{ii}}(2)$ is
different\cite{KL}). The BA equations are now:
\begin{equation}
\xi _a^{2N}=\prod_{b=1,b\neq a}^r\;\frac{b(\xi _a^{-1},\xi _b)}{b(\xi _a,\xi
_b)},\;\;a=1,2,  \label{free}
\end{equation}
where
\begin{equation}
b(\xi _a,\xi _b)=\frac{\xi _b}{\xi _a}[\xi _b+\xi _a^{-1}-2\Delta -E_{ab}][%
\xi _b^{-1}+\xi _a^{-1}-2\Delta -E_{ab}].
\end{equation}
and
\begin{equation}
E_{ab}=2\sum_l^{\prime }q^{-<e_l,\rho >}+\Gamma (\xi _a)\Gamma ({\xi }%
_a^{-1})+\Gamma (\xi _b)\Gamma ({\xi }_b^{-1}).
\end{equation}

The only other block is $\epsilon_1=\epsilon_2=0, E=0$. It is again highly
degenerate with ${\cal F}_{b_{ii}}(2),\xi_1,\xi_2$ as free parameters%
\footnote{%
Actually there is one more free parameter, called $\alpha_5$ in ref.\cite{KL}%
.}.

Thus all models have {\em spectra equivalent to the one of the XXZ model}.

\section{Conclusion}

We obtained the spectra of quantum spin chain models, arising as
representations of the Temperley-Lieb algebra associated with  quantum
groups. The tool is a modified version of the coordinate Bethe Ansatz,
since the simpler algebraic Bethe Ansatz is not immediately available for
these models. We find that all models have equivalent spectra, i.e. they
differ at most in their degeneracies. The energy eigenvalues are given by
\begin{equation}
        E=\sum_{n=1}^p( \eps_n
        ( -2*\Delta +\Gamma (\xi_n)\Gamma ({\xi }_n^{-1})),
\end{equation}
where $-2\Delta=\sum_l^{\prime }q^{-<e_l,\rho >}$ and the rapidities ${\xi }_n$
are solutions of the BA equations.

In the sector $r$ we may have $p$
pseudoparticles $N_{\omega ^{*}-1},N_{\omega ^{*}-1},\cdots ,N_{-\omega
^{*}+1}$ impurities of the type $(\omega ^{*}-1),$ $(\omega ^{*}-2),\cdots
,(-\omega ^{*}+1)$, respectively, such that
\begin{equation}
N_{\omega ^{*}-1}+2N_{\omega ^{*}-1}+\cdots +(2\omega ^{*}-1)N_{-\omega
^{*}+1}=r-2\omega ^{*}p.
  \label{sect}
\end{equation}
Here $\omega ^{*}=\omega $ for $A_1$ and $B_n$ and $\omega ^{*}=\omega -1/2$
for $C_n$ and $D_n$.

For example, for periodic boundary conditions\footnote{We don't list the
equations for free boundary conditions, since they are identical to those of
ref.\cite{KL}.}, the total rapidity
 $\xi =\xi _1\xi _2\ldots \xi _p\xi_{imp},\,\xi _{imp}=\xi _{p+1}\xi
_{p+2}\ldots \xi _{r-\sigma }$
 obeys
$\xi^N=1$ and the BA equations for $E\neq 0$ are:
\begin{equation}
\xi _a^N\xi _{imp}^2=\prod_{b=1,b\neq a}^{r-\sigma }-\frac{\xi _a}{\xi _b}%
\frac{(1+(\xi _a\xi _b)^{-1}\xi _a-\epsilon _{ab}}{(1+(\xi _a\xi _b)^{-1}\xi
_b-\epsilon _{ab}}\;,
\end{equation}
where $\epsilon _{ab}=E_{ab}+2\Delta $ and $\sigma $ can be fixed for each
allowed case from the equations (\ref{sect}).The Bethe eigenstates are
highest weight states of the quantum group, except for the states with
energy $E=0$, for which this is not always true.  in the sector $r$ with $p$
impurities and $E\neq 0$, the total rapidity
$\xi=\xi_1\xi_2\ldots\xi_p\xi_{imp},\,\xi_{imp}=\xi_{p+1}\xi_{p+2}\ldots\xi_{r-p}$ obeys $\xi^N=1$ and the BA equations are:
\beq
  \xi_a^N \xi_{imp}^2 = \prod_{b=1,b\neq a} -\frac{\xi_a}{\xi_b}
         \frac{(1+(\xi_a\xi_b)^{-1}\xi_a - \eps}
              {(1+(\xi_a\xi_b)^{-1}\xi_b - \eps}\;,
\eeq
where $\eps=E+2\Delta$. The Bethe eigenstates are highest
weight states of the quantum group, except for the states with energy $E=0$,
for which this is not always true. These results are expected, but as far as we
know, unprooven using only Temperley-Lieb algebraic statements as input.

{\bf \flushleft  Acknowledgments}

It is R.K.'s pleasure to thank C.Callan and W. Bialek for their support.

%

\end{document}